\documentclass[11pt,a4paper]{article}

\usepackage[utf8]{inputenc}
\usepackage[T1]{fontenc}
\usepackage{amsmath,amssymb}
\usepackage{graphicx}
\usepackage{booktabs}
\usepackage{hyperref}
\usepackage{url}
\usepackage{natbib}
\usepackage[margin=2.5cm]{geometry}
\usepackage{tabularx}
\usepackage{array}
\usepackage{caption}
\usepackage{float}
\usepackage{microtype}
\usepackage{authblk}

\usepackage{doi}

\hypersetup{
    colorlinks=true,
    linkcolor=blue,
    citecolor=blue,
    urlcolor=blue
}
\title{Where does a Rust speedup come from? Language and algorithm effects in a sliding window threat scorer}

\author[1]{Nikolaos D.~Tantaroudas\thanks{Corresponding author, \texttt{nikolaos.tantaroudas@iccs.gr}}}
\author[2]{Ilias Karachalios}
\author[3]{Andrew J.~McCracken}

\affil[1]{Institute of Communication and Computer Systems (ICCS), National Technical University of Athens, Iroon Polytechneiou 9, 15773 Zografou, Athens, Greece}
\affil[2]{University of Thessaly, 413 34, Larissa, Greece}
\affil[3]{DASKALOS APPS SAS, 183 Rue de l'Abb\'{e} Griffon, 01960 P\'{e}ronnas, France}

\date{}

\begin{document}

\maketitle

\begin{abstract}
Rewriting a hot path from Python into Rust is a common way to speed up security analytics, and large speedups are routinely reported. A rewrite usually changes the language and the algorithm at once, so a single factor can credit the language with a gain that comes from a better algorithm. We study this on a sliding window threat scorer modelled on the traffic light calculator of the SentinelSphere platform. Five implementations, three in Python and two in Rust, produce bit identical scores, confirmed by a shared checksum, and we time them from 100 to one million events in a bounded and a burst regime. The language alone contributes between about 4 and 29 times. Replacing a per event rescan of the window by an incremental update contributes more than 11,000 times in the burst regime, so the end to end factor reaches about 195,000 times when the window keeps filling and levels off near 13,000 times when it does not. A power law fit to the timings reported for the original rewrite gives growth exponents of 1.85 for Python and 0.88 for Rust, the signature of an algorithmic difference. Performance claims for security tooling should therefore report the language and algorithm contributions separately.
\end{abstract}


\section{Introduction}
\label{sec:introduction}

Security pipelines increasingly process event streams in near real time, and a frequent engineering step is to rewrite a Python hot path in a systems language such as Rust. The resulting speedups are often quoted as one multiplicative factor and attributed to the language. The SentinelSphere platform~\citep{tantaroudas2026ore,tantaroudas2026arxiv} reports such a rewrite of its anomaly scoring component, with an average speedup of 5.6 times in steady operation and factors from 4.3 to 326 times for batches of 1,000 to 100,000 events. A factor that keeps rising with the batch size is unusual for a pure language effect, because native compilation, the absence of an interpreter and contiguous memory give a roughly constant gain, whereas a factor that grows with the input is the signature of a change in algorithmic complexity.

We therefore ask how much of such a speedup is due to the language and how much to a change in the algorithm made at the same time. We hold the scoring semantics fixed and separate the two axes by implementing a naive and an incremental version in both languages, measuring repeatedly and summarising robustly as recommended for performance evaluation~\citep{georges2007,kalibera2013}. Our contribution is a clean measurement and a reusable harness rather than a new detector.

\section{Method}
\label{sec:method}

The workload is a sliding window scorer that mirrors the SentinelSphere traffic light calculator~\citep{tantaroudas2026ore,tantaroudas2026arxiv}. For every event it keeps a window of the last 300 seconds, computes a time decayed sum of event severities and multiplies it by five contextual factors, the event rate in the last minute, the number of temporal clusters, the concentration of source addresses, the diversity of attack types and the duration of sustained activity, before capping the score at 100.

Three Python implementations, a naive rescan that reads the window by index, a naive rescan that iterates over it and an incremental version that maintains running counters, and two Rust implementations, a naive rescan and an incremental version, compute the same scores. A checksum over the full sequence of capped and uncapped scores agreed across all five implementations on identical input, and a 64 bit linear congruential generator with a fixed seed produces bit identical event streams in both languages. In the bounded regime the inter arrival times keep the window at a roughly constant size, as in steady operation, and in the burst regime events arrive almost together so that the window grows with the input, as in a flood or a backlog replay. We time the scoring loop only, take the median of ten repetitions for the fast implementations and fewer repetitions for the slowest ones, and run everything on two cores of an Intel Xeon processor at 2.8\,GHz with Python 3.13 and a Rust release build with link time optimisation.

\section{Results}
\label{sec:results}

\begin{figure}[H]
    \centering
    \includegraphics[width=\textwidth]{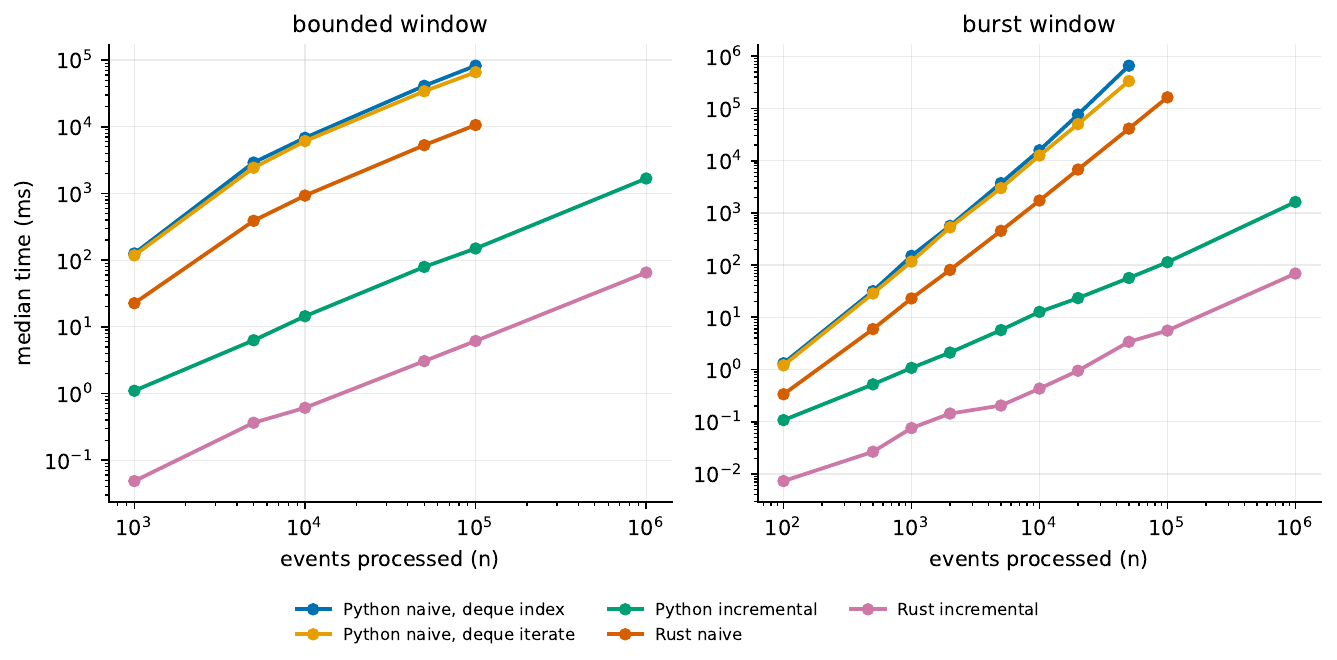}
    \caption{Median scoring time against the number of events for the five implementations, on logarithmic axes, in the bounded regime (left) and the burst regime (right).}
    \label{fig:time}
\end{figure}

The score checksums agreed across all implementations and sizes. In the burst regime the naive implementations grow with exponents close to two, 2.1 for the Python index version, 2.0 for the iterating version and 1.9 for Rust, while both incremental versions grow with an exponent close to one (Figure~\ref{fig:time}). In the bounded regime all implementations grow roughly linearly once the window is full, with the naive versions paying a constant multiple of the window size for every event.

\begin{figure}[H]
    \centering
    \includegraphics[width=0.72\textwidth]{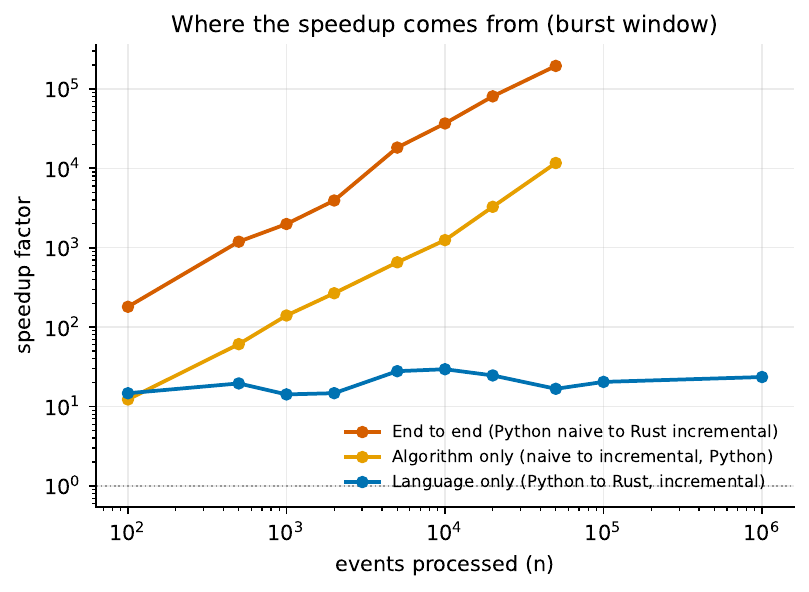}
    \caption{End to end, algorithm only and language only factors against the number of events in the burst regime. The language factor stays nearly flat while the other two grow with the input.}
    \label{fig:decomposition}
\end{figure}

Table~\ref{tab:decomposition} decomposes the end to end factor. The language factor, Python over Rust on the same algorithm, ranges from about 14 to 29 times for the incremental algorithm and from about 4 to 16 times for the naive one, across all sizes and regimes. The algorithm factor, naive over incremental within Python, rises from 115 times at 1,000 events to 551 times at 100,000 events in the bounded regime, where it levels off because the window has stopped growing, and reaches 11,652 times at 50,000 events in the burst regime. The end to end factor is the product of the two and only its algorithmic part grows with the input (Figure~\ref{fig:decomposition}).

\begin{table}[H]
\centering
\caption{Decomposition of the end to end speedup into a language factor and an algorithm factor.}
\label{tab:decomposition}
\small
\begin{tabularx}{\textwidth}{l r >{\raggedleft\arraybackslash}X >{\raggedleft\arraybackslash}X >{\raggedleft\arraybackslash}X}
\toprule
\textbf{Regime} & \textbf{Events} & \textbf{Language factor, incremental} & \textbf{Algorithm factor, Python} & \textbf{End to end factor} \\
\midrule
bounded & 1,000 & 22.6 & 114.6 & 2,591 \\
bounded & 10,000 & 23.6 & 473.5 & 11,187 \\
bounded & 100,000 & 24.4 & 551.5 & 13,442 \\
burst & 10,000 & 29.5 & 1,245.9 & 36,724 \\
burst & 50,000 & 16.7 & 11,652.4 & 194,714 \\
\bottomrule
\end{tabularx}
\end{table}

\section{Discussion and conclusions}
\label{sec:discussion}

The batch factor reported for the original rewrite climbs with the batch size in the same way. Fitting a power law to the batch timings that the article lists for each batch size~\citep{tantaroudas2026ore} gives exponents of about 1.85 for Python and 0.88 for Rust, the contrast between near quadratic and near linear growth that separates our naive and incremental implementations. The per event cycle counts quoted there, about 314,000 for Python against 6,375 for Rust, imply a ratio near 50 rather than the headline factor, and the steady state average of 5.6 times lies within the range we measure for the language alone on the naive algorithm in the bounded regime, about 5.5 to 8 times. Most of the large batch benefit was therefore available in Python by fixing the algorithm, with the language adding a bounded factor on top.

A rewrite that changes both the language and the algorithm should report the language only factor and the algorithm only factor together with the input sizes, because an inefficient baseline can otherwise inflate a reported number, a challenge also identified for machine learning in security~\citep{arp2024}. Separating the two costs little, since it only requires the original algorithm in the new language or the new algorithm in the original one. Our numbers come from one workload on one processor and should be read as the shape of the effect rather than as constants, and energy, which also depends on the language~\citep{pereira2021}, was not measured.

\section*{Data and code availability}

The Python and Rust implementations, the benchmark harness, the raw timings and the analysis scripts can be provided as ancillary files with this preprint if requested.

\section*{Acknowledgements}

This work was supported by the European Union's Horizon Europe programme under Grant Agreement No.\ 101119681 (ResilMesh). Views and opinions expressed are those of the authors only and do not necessarily reflect those of the European Union or the granting authority. Neither the European Union nor the granting authority can be held responsible for them. The authors used a generative AI assistant to help implement the harness, run the measurements and draft the text, and they reviewed and verified all content.

\end{document}